\documentclass[conference]{IEEEtran}
\usepackage[T1]{fontenc}
\usepackage[utf8]{inputenc}
\usepackage[english]{babel}
\usepackage{amsmath}
\usepackage{amsthm}
\usepackage{amsfonts}
\usepackage{colonequals}
\usepackage[cmintegrals]{newtxmath}
\usepackage{mathtools}
\usepackage{thmtools}
\usepackage{stmaryrd}
\usepackage{bm}
\usepackage{siunitx}
\usepackage{comment}
\usepackage{microtype}
\usepackage{fnpct}
\usepackage{fontawesome5}
\usepackage{csquotes}
\usepackage[inline]{enumitem}
\usepackage{tabularx}
\usepackage{booktabs}
\usepackage{textcomp}
\usepackage[hidelinks]{hyperref}
\usepackage[capitalize,nameinlink]{cleveref}
\usepackage{xcolor}
\usepackage{graphicx}
\usepackage[inkscapelatex=false]{svg}
\usepackage{wrapfig}
\usepackage{float}
\usepackage{placeins}
\usepackage{xspace}
\usepackage[noEnd]{algpseudocodex}
\usepackage[
    backend=biber,
    style=ieee,
    sortlocale=en_US,
    sortcites=true,
    url=true,
    eprint=true,
    giveninits=false,
    dashed=false,
    minnames=1,
    maxnames=5
]{biblatex}
\AtEveryBibitem{%
    \clearname{editor}%
    \clearfield{series}%
    \clearfield{isbn}%
    \clearfield{issn}%
    \clearfield{day}%
    \clearfield{month}%
    \clearfield{place}%
    \clearlist{location}%
    \clearfield{pages}%
    \clearfield{volume}%
    \clearfield{number}%
}
\DeclareSourcemap{
    \maps[datatype=bibtex]{
        \map{
            \step[fieldsource=doi, match=\regexp{.+/arXiv\..+}, final]
            \step[fieldsource=eprint, match=\regexp{.+}, final]
            \step[fieldset=doi, null]
        }
        \map{
            \step[fieldsource=doi, match=\regexp{.+}, final]
            \step[fieldset=eprint, null]
            \step[fieldset=archiveprefix, null]
            \step[fieldset=eprinttype, null]
        }
    }
}
\setlist[enumerate]{label=(\arabic*)}

\makeatletter
\long\def\@makecaption#1#2{%
    \ifx\@captype\@IEEEtablestring%
    \footnotesize\bgroup\par\centering\@IEEEtabletopskipstrut{\normalfont\footnotesize {#1.}\nobreakspace\scshape #2}\par\addvspace{0.5\baselineskip}\egroup%
    \@IEEEtablecaptionsepspace
    \else
        \@IEEEfigurecaptionsepspace
        \setbox\@tempboxa\hbox{\normalfont\footnotesize {#1.}\nobreakspace #2}%
        \ifdim \wd\@tempboxa >\hsize%
        \setbox\@tempboxa\hbox{\normalfont\footnotesize {#1.}\nobreakspace}%
        \parbox[t]{\hsize}{\normalfont\footnotesize \noindent\unhbox\@tempboxa#2}%
        \else%
            \hbox to\hsize{\normalfont\footnotesize\hfil\box\@tempboxa\hfil}%
        \fi\fi}
\makeatother
\makeatletter
\let\MYcaption\@makecaption
\makeatother
\usepackage{subcaption}
\makeatletter
\let\@makecaption\MYcaption
\makeatother
\svgpath{ {figures/} }
\faStyle{regular}

\declaretheoremstyle[
    bodyfont=\itshape,%
]{example-style}
\declaretheorem[
    name=Example,%
    style=example-style,%
    numbered=unless unique,%
]{example}

\crefname{section}{Sec.}{Sec.}
\Crefname{section}{Sec.}{Sec.}
\crefname{example}{Ex.}{Ex.}
\Crefname{example}{Ex.}{Ex.}
\newcolumntype{R}{>{\raggedleft\arraybackslash}X}
\newcolumntype{C}{>{\centering\arraybackslash}X}
\definecolor{TUM_blue}{RGB}{0,101,189}
\colorlet{TUM_black}{black}
\colorlet{TUM_white}{white}
\definecolor{TUM_darkblue}{RGB}{0,82,147}
\colorlet{TUM_darkblue100}{TUM_darkblue}
\colorlet{TUM_darkblue80}{TUM_darkblue100!80}
\colorlet{TUM_darkblue50}{TUM_darkblue100!50}
\colorlet{TUM_darkblue20}{TUM_darkblue100!20}
\definecolor{TUM_verydarkblue}{RGB}{0,51,89}
\colorlet{TUM_verydarkblue100}{TUM_verydarkblue}
\colorlet{TUM_verydarkblue80}{TUM_verydarkblue100!80}
\colorlet{TUM_verydarkblue50}{TUM_verydarkblue100!50}
\colorlet{TUM_verydarkblue20}{TUM_verydarkblue100!20}
\colorlet{TUM_darkgrey}{TUM_black!80}
\colorlet{TUM_grey}{TUM_black!50}
\colorlet{TUM_lightgrey}{TUM_black!20}
\definecolor{TUM_beige}{RGB}{218,215,203}
\definecolor{TUM_orange}{RGB}{227,114,34}
\definecolor{TUM_green}{RGB}{162,173,0}
\definecolor{TUM_verylightblue}{RGB}{152,198,234}
\definecolor{TUM_lightblue}{RGB}{100,160,200}
\newcommand{\ie}{i.\,e.\nolinebreak\@\xspace}

\newcommand{\eg}{e.\,g.\nolinebreak\@\xspace}

\renewcommand{\phi}{\varphi}
\newcommand{\defeq}{\colonequals}
\newcommand{\abs}[1]{\left|#1\right|}
\newcommand{\set}[1]{\left\{#1\right\}}

\title{Routing-Aware Placement\\for Zoned Neutral Atom-based Quantum Computing}
\author{
    \IEEEauthorblockN{
        Yannick Stade\IEEEauthorrefmark{1},
        Wan-Hsuan Lin\IEEEauthorrefmark{4},
        Jason Cong\IEEEauthorrefmark{4},
        and Robert Wille\IEEEauthorrefmark{1}\IEEEauthorrefmark{2}\IEEEauthorrefmark{3}
    }
    \IEEEauthorblockA{\IEEEauthorrefmark{1}%
    Chair for Design Automation,
        Technical University of Munich,
        Munich, Germany
    }
    \IEEEauthorblockA{\IEEEauthorrefmark{2}%
    Munich Quantum Software Company GmbH,
        Garching near Munich, Germany
    }
    \IEEEauthorblockA{\IEEEauthorrefmark{3}%
    Software Competence Center Hagenberg GmbH,
        Hagenberg, Austria
    }
    \IEEEauthorblockA{\IEEEauthorrefmark{4}%
    Computer Science Department,
        University of California,
        Los Angeles, USA
    }
    yannick.stade@tum.de, %
    wanhsuanlin@ucla.edu, %
    cong@cs.ucla.edu, %
    robert.wille@tum.de\\
    \href{https://www.cda.cit.tum.de/research/quantum}{www.cda.cit.tum.de/research/quantum}
}
\hypersetup{ %
    pdftitle={Routing-Aware Placement for Zoned Neutral Atom-based Quantum Computing},
    pdfsubject={IEEE International Conference on Computer-Aided Design 2025},
    pdfauthor={
        Yannick Stade,
        Wan-Hsuan Lin,
        Jason Cong,
        Robert Wille
    }
}
\begin{document}
    \maketitle

    \begin{abstract}
        Quantum computing promises to solve previously intractable problems, with neutral atoms emerging as a promising technology.
        Zoned neutral atom architectures allow for immense parallelism and higher coherence times by shielding idling atoms from interference with laser beams.
        However, in addition to hardware, successful quantum computation requires sophisticated software support, particularly compilers that optimize quantum algorithms for hardware execution.
        In the compilation flow for zoned neutral atom architectures, the effective interplay of the placement and routing stages decides the overhead caused by rearranging the atoms during the quantum computation.
        Sub-optimal placements can lead to unnecessary serialization of the rearrangements in the subsequent routing stage.
        Despite this, all existing compilers treat placement and routing independently thus far---focusing solely on minimizing travel distances.
        This work introduces the first \mbox{routing-aware} placement method to address this shortcoming.
        It groups compatible movements into parallel rearrangement steps to minimize both rearrangement steps and travel distances.
        The implementation utilizing the A* algorithm reduces the rearrangement time by 17\% on average and by 49\% in the best case compared to the \mbox{state-of-the-art}.
        The complete code is publicly available in open-source as part of the \emph{Munich Quantum Toolkit}~(MQT) at \url{https://github.com/munich-quantum-toolkit/qmap}.
    \end{abstract}


    \section{Introduction}\label{sec:introduction}
    Quantum computing promises to solve problems deemed intractable before~\cite{preskillQuantumComputingNISQ2018}.
    Many different technologies are being explored---with neutral atoms being a promising candidate~\cite{bluvsteinLogicalQuantumProcessor2023}.
    Here, the state of the atoms---representing the qubits---is manipulated by laser beams.
    Those laser beams can illuminate many atoms simultaneously, allowing for immense parallelism.
    In particular, zoned neutral atom architectures~\cite{bluvsteinLogicalQuantumProcessor2023} emerged as a promising candidate for quantum computing.
    Experiments on those architectures have demonstrated basic quantum computations with up to 280 physical qubits~\cite{bluvsteinLogicalQuantumProcessor2023}.
    These zoned architectures allow the shielding of idling atoms from interference with laser beams used to perform the operations.
    Shielding the idling atoms leads to significantly higher coherence times than monolithic architectures~\mbox{\cite{stadeOptimalStatePreparation2024,linReuseAwareCompilationZoned2024}}.

    However, just having the hardware is not sufficient to perform quantum computations.
    Software is key to exploiting the hardware's full potential.
    In particular, compilers automate the translation of quantum algorithms into an optimized sequence of instructions that can be executed on the hardware.
    For other quantum technologies already a broad spectrum of compiler methods have been proposed, \eg, for superconducting qubits~\mbox{\cite{burgholzerLimitingSearchSpace2022,fuEffectiveEfficientQubit2023,shaikOptimalLayoutSynthesis2023}, ion traps~\cite{schmaleBackendCompilerPhases2022,kreppelQuantumCircuitCompiler2023,schoenbergerCyclebasedShuttlingTrappedIon2024}}, etc.
    In contrast, the software support for zoned neutral atom architectures~\mbox{\cite{stadeAbstractModelEfficient2024,linReuseAwareCompilationZoned2024}} remains rather limited.

    More precisely, the compilation is separated into several stages---with \emph{placement} and \emph{routing} being particularly important as part of the layout synthesis.
    Their interplay decides how efficiently the atoms are rearranged during the quantum computation.
    Obviously, the placement affects the subsequent routing: A bad placement leads to bad routing.
    Unfortunately, all available compilers handle the placement and routing as two completely independent stages and purely focus on minimizing atoms' travel distance---leaving substantial room for improvement (as motivated in more detail later in~\cref{sec:problem}).

    This work addresses this problem by proposing the first \emph{\mbox{routing-aware} placement} method for zoned neutral atom architectures to facilitate efficient routing.
    This method groups compatible movements that can be executed in parallel into rearrangement steps.
    The objective of \mbox{routing-aware} placement is to minimize not only the travel distances of the atoms but also the number of rearrangement steps.
    Given the vast search space of possible placements, sophisticated exploration methods are essential.
    Therefore, we propose a solution based on the A* algorithm~\cite{hartFormalBasisHeuristic1968}, complemented by a dedicated data structure for efficient and accurate cost anticipation.

    The evaluation shows that the proposed \mbox{routing-aware} placement significantly reduces the number of rearrangement steps, which eventually reduces the time required to perform the rearrangements by up to 49\% in the best case and 17\% on average.
    The implementation of the proposed approach is publicly available as part of the \emph{Munich Quantum Toolkit}~(MQT,~\cite{willeMQTHandbookSummary2024}) at \url{https://github.com/munich-quantum-toolkit/qmap}.

    \section{Background}\label{sec:background}

    This section briefly revisits the fundamentals of neutral atom-based quantum computing to keep this paper self-contained.
    \Cref{sec:zoned architecture} reviews zoned neutral atom architectures,
    and \cref{sec:zoned compilers} provides an overview of existing works on the compilation for neutral atoms.

    \subsection{Zoned Neutral Atom Architectures}\label{sec:zoned architecture}
    In quantum computing based on neutral atoms~\mbox{\cite{bluvsteinQuantumProcessorBased2022,schmidComputationalCapabilitiesCompiler2024}}, qubits are encoded in the electronic states of individual neutral atoms such as Rubidium~(Rb), Strontium~(Sr), or Ytterbium~(Yb).
    Those atoms are confined in optical tweezers and cooled with lasers to their motional ground state~\mbox{\cite{gygerContinuousOperationLargescale2024,barredoAtombyatomAssemblerDefectfree2016}}.
    \mbox{One-qubit} operations, such as rotational operations, are realized through state transitions driven by global and local lasers~\mbox{\cite{saffmanQuantumComputingNeutral2019,bluvsteinQuantumProcessorBased2022}}.
    \mbox{Two-qubit} operations, such as CZ operations, are realized by global Rydberg beams~\mbox{\cite{everedHighfidelityParallelEntangling2023,giudiciFastEntanglingGates2024}}.
    The Rydberg blockade mechanism ensures that only the qubits within the interaction radius of each other interact in the current technology.
    However, isolated atoms still experience an imperfect identity operation as they are still excited to the Rydberg state, leading to Rydberg decay---a significant source of errors~\cite{everedHighfidelityParallelEntangling2023}.

    \begin{figure}[tp]
        \vspace{-7pt}
        \centering
        \includegraphics[width=\linewidth]{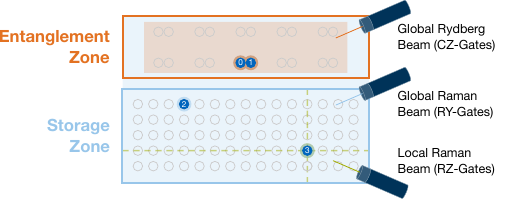}\\
        \vspace{-10pt}
        \caption{Schematic of a zoned neutral atom architecture illustrating global and local operations on atoms (blue).}
        \label{fig:zones}
        \vspace{-10pt}
    \end{figure}

    To mitigate those errors, zoned architectures perform operations in designated spatially separated zones~\cite{bluvsteinLogicalQuantumProcessor2023}, such as the \emph{entanglement zone} and the \emph{storage zone}, as shown in~\cref{fig:zones}.
    The entanglement zone features spatially separated pairs of traps such that atoms in the same pair can interact with each other but not with atoms in other traps.
    The Rydberg beam only affects atoms in the entanglement zone, and multiple CZ operations can be performed in parallel.
    During \mbox{two-qubit} operations, atoms in the storage zone are shielded---leading to significantly higher coherence times~\cite{bluvsteinLogicalQuantumProcessor2023}.
    Without the need to keep a separation between atoms for unwanted interactions, the storage zone offers many densely packed traps to store atoms.
    Each trap is realized as a static optical tweezer created, \eg, by a \emph{Spatial Light Modulator}~(SLM,~\cite{bluvsteinQuantumProcessorBased2022}) and holds up to one atom (aka qubit).

    The usual course of execution of \mbox{two-qubit} operations is to move qubits from the storage zone to the entanglement zone, perform the operation, and move them back to the storage zone.
    Those rearrangements are realized by an additional kind of adjustable optical tweezers controlled by \emph{Acousto-Optic Deflectors}~(AODs,~\cite{bluvsteinQuantumProcessorBased2022}).
    These AODs can pick up atoms from static traps, move them via stretching the AOD columns and rows, and drop them in other static traps, constituting one \emph{rearrangement step}.
    Transferring between traps may cause atom loss~\cite{bluvsteinLogicalQuantumProcessor2023}, which can be solved by atom reload in the future~\cite{gygerContinuousOperationLargescale2024}.
    On the other hand, atom movement does not incur errors in the qubits if the speed is below a certain threshold.

\begin{figure}[pb]
        \vspace{-8pt}
        \centering
        \includegraphics[width=\linewidth]{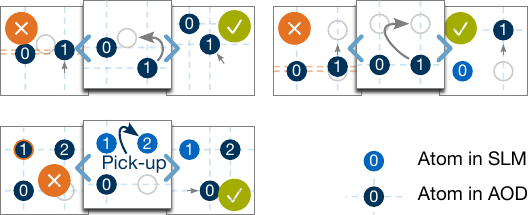}\\
        \raggedright\vspace{-62pt}
        \begin{subfigure}{118pt}
            \caption{Non-crossing constraint}
            \label{fig:constraint:non-crossing}
        \end{subfigure}
        \hspace{6pt}
        \begin{subfigure}{118pt}
            \caption{Preservation constraint}
            \label{fig:constraint:preservation}
        \end{subfigure}\\
        \vspace{42pt}
        \begin{subfigure}{118pt}
            \caption{Ghost-spot constraint}
            \label{fig:constraint:ghost-spots}
        \end{subfigure}
        \vspace{-6pt}
        \caption{
            Rearrangement Constraints: The middle frame of each sub-figure shows the intended rearrangement, the left one a violation, and the right a possible workaround.
        }
        \label{fig:constraints}
    \end{figure}

    Multiple atom movements can be performed in parallel, referred to as one \emph{rearrangement}.
    However, in doing so, the following \emph{rearrangement constraints} (as specified in~\cite{stadeAbstractModelEfficient2024} and illustrated in \cref{fig:constraints}) must be obeyed:
    \begin{itemize}
        \item \emph{Non-Crossing}: During one rearrangement step, the rows and columns of AOD traps must not cross each other to prevent atom loss and heating~\cite{stadeAbstractModelEfficient2024}, cf.~\cref{fig:constraint:non-crossing}.
        \item \emph{Preservation}: Additionally, qubits starting in the same row or column must stay in the same row or column, respectively, cf.~\cref{fig:constraint:preservation}.
        \item \emph{Ghost-Spots}: When transferring atoms from and to AOD traps, atoms at all grid points of the rows and columns are affected, cf.~\cref{fig:constraint:ghost-spots}.
    \end{itemize}
    Consequently, the rearrangement of atoms in preparation of a layer of \mbox{two-qubit} operations may incur multiple rearrangement steps.
    However, since the state of the atoms decoheres over time~\cite{bluvsteinLogicalQuantumProcessor2023}, the duration of the quantum computation must be minimized to achieve high fidelity.
    Thus, the rearrangement constraints are central to the compilation and significantly contribute to its complexity.

    \subsection{Overview of Corresponding Compilers}\label{sec:zoned compilers}
    Compiling quantum circuits with many quantum gates to different kinds of neutral atom architectures is an active research field.
    Early developments focused on an architecture with individually addressable entangling gates and SWAP gates to route qubits on a grid of static traps~\mbox{\cite{bakerExploitingLongDistanceInteractions2021,patelGeyserCompilationFramework2022,schmidHybridCircuitMapping2024}}.
    However, the optical setup required for individually addressable entangling gates could not reach the same fidelity as zoned architectures yet (92.5\% vs. 99.5\% fidelity~\mbox{\cite{grahamMultiqubitEntanglementAlgorithms2022,everedHighfidelityParallelEntangling2023}}).
    A different line of research focuses on architectures capable of rearranging atoms with AODs but still with a monolithic design, \ie, one zone with a global Rydberg beam~\mbox{\cite{constantinidesOptimalRoutingProtocols2024,huangDasAtomDivideandShuttleAtom2024,ludmirPARALLAXCompilerNeutral2024,nottinghamDecomposingRoutingQuantum2023,patelGRAPHINEEnhancedNeutral2023,silverQomposeTechniqueSelect2024,tanCompilationDynamicallyFieldProgrammable2025,tanCompilingQuantumCircuits2024,tanDepthOptimalAddressing2D2024,tanQubitMappingReconfigurable2022,wangAtomiqueQuantumCompiler2024,wangQPilotFieldProgrammable2024}}.
    So far, only a few compilers consider zoned architectures.

    \begin{figure*}[tp]
        \centering
        \includegraphics[width=\linewidth]{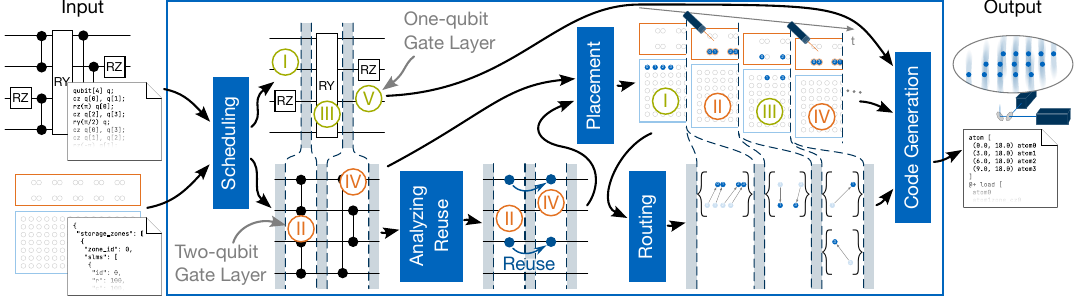}\\
        \vspace{-6pt}
        \caption{
            The common compilation flow for zoned neutral atom architectures.
            It takes a quantum circuit and the zoned neutral atom architecture specification as input and generates a sequence of target-specific instructions.
        }
        \label{fig:software architecture}
        \vspace{-10pt}
    \end{figure*}

    In fact, NALAC just recently proposed in~\cite{stadeAbstractModelEfficient2024} initiated this line of research.
    It reduces the time-costly trap transfers by keeping one set of atoms in AODs while sliding past the other set of atoms and performing multiple CZ-gates in one go.
    The tool, proposed in~\cite{stadeOptimalStatePreparation2024}, generates optimal schedules for state preparation circuits used for quantum error correction.
    Its evaluation demonstrates clearly that shielding idling qubits in the storage zone is crucial to achieving high fidelity.
    A fact that the tools ZAC~\cite{linReuseAwareCompilationZoned2024} and ZAP~\cite{huangZAPZonedArchitecture2024} take advantage of.
    In addition, they detect reuse opportunities, \ie, they keep qubits in the entanglement zone if they are involved in consecutive \mbox{two-qubit} gates.
    Besides those general-purpose compilers, Artic~\cite{deckerArcticFieldProgrammable2024} was proposed to tackle one specific architecture.
    Mantra~\cite{jangQubitMovementOptimizedProgram2025} targets the architecture, which can only execute \mbox{one-qubit} gates in the storage zone.
    This tool defines rewrite rules to replace common patterns in quantum circuits with ones better suited for neutral atom architectures.
    However, all these methods are too specific for compiling general quantum circuits or generate an unnecessary large rearrangement overhead.

    \section[Motivation: Untapped Potential in Existing Compilers]{Motivation:\\Untapped Potential in Existing Compilers}\label{sec:problem}
    Following the established compilation flow, the compilation of quantum computations
    for zoned neutral atom architectures is divided into several stages.
    The fidelity of the resulting quantum computation depends mutually on the result of all stages.
    This section reviews those stages and then unveils substantial potential missed in available compilers.

    \subsection{Compilation Flow}\label{sec:compilation flow}
    A compiler receives
    \begin{enumerate*}[label=(\arabic*)]
        \item a quantum circuit to execute and
        \item a zoned neutral atom architecture specification as input.
    \end{enumerate*}
    It transforms the quantum circuit into a sequence of target-specific instructions that align with the constraints of the architecture.
    To this end, the flow is usually separated into several stages as illustrated in~\cref{fig:software architecture} and briefly reviewed in the following:

    \paragraph*{Scheduling} This stage schedules all gates of the input circuit into two types of \emph{layer}: \mbox{One-qubit} gate and \mbox{two-qubit} gate layers.
    All \mbox{two-qubit} gates in one layer must be executable in parallel, \ie, they must not share qubits.
    The resulting schedule consists of a sequence of alternating \mbox{one-qubit} and \mbox{two-qubit} gate layers---some \mbox{one-qubit} gate layers may be empty, though.
    The default setting for this stage is an \mbox{as-soon-as-possible} scheduling as, \eg, in~\cite{linReuseAwareCompilationZoned2024} that puts every gate in the earliest possible layer.

    \paragraph*{Analyzing Reuse} This stage is optional but very beneficial and, \eg, used by compilers such as~\cite{linReuseAwareCompilationZoned2024}.
    It detects reuse opportunities:
    When an atom takes part in two consecutive \mbox{two-qubit} gate layers, it can remain in the entanglement zone, \ie, it can be \emph{reused} without the need to move it back to the storage zone.
    This significantly improves the overall fidelity because it reduces the number of necessary trap transfers.

    \paragraph*{Placement} This stage determines the location of every atom in each layer.
    It takes the set of independent \mbox{two-qubit} gates per \mbox{two-qubit} gate layer from the schedule.
    Based on this input, it produces the resulting placement as output.
    In every \mbox{two-qubit} gate layer, the atoms belonging to one gate must be placed in a pair of traps in the entanglement zone, cf.~\cref{sec:zoned architecture}.
    All remaining atoms should be placed in the storage zone to shield them from the Rydberg beam and avoid decoherence due to the Rydberg decay.

    \paragraph*{Routing} The routing stage takes the placement and determines the necessary rearrangements of the atoms to transition from the placement of one layer to the next.
    The rearrangement constraints (cf.~\cref{sec:zoned architecture}) must be obeyed during this process.
    These constraints may necessitate multiple rearrangement steps to transition from one layer to the next.

    \paragraph*{Code Generation} The final stage combines the results from the previous stages.
    It incorporates the schedule, the placement, and the routing determined before and generates a sequence of target-specific instructions.
    The output is a program to be executed on a zoned neutral atom quantum computer.

    \subsection{Potential for Improvement}\label{sec:potential}

    As reviewed above, the placement significantly impacts the routing stage during the compilation flow.
    It directly influences the level of parallelism one can achieve in the routing stage and, by this, the rearrangement time during the quantum computation.
    However, placement methods such as, \eg, proposed in~\cite{linReuseAwareCompilationZoned2024,huangZAPZonedArchitecture2024}, heavily rely their placement on minimizing the accumulated travel distance of atoms.
    While this, at first glance, seems like an appropriate heuristic, it may make routing afterward considerably harder.

    \begin{figure}[bp]
        \vspace{-8pt}
        \centering
        \includegraphics[width=\linewidth]{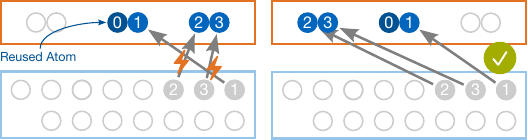}\\
        \raggedright\vspace{-18pt}\hspace{4pt}%
        \begin{subfigure}{127pt}
            \caption{\raggedright}
            \label{fig:placement:bad}
        \end{subfigure}
        \begin{subfigure}{100pt}
            \caption{\raggedright}
            \label{fig:placement:good}
        \end{subfigure}
        \vspace{-4pt}
        \caption{
            The \mbox{routing-agnostic} placement (\subref{fig:placement:bad}) requires two whereas the routing aware-placement (\subref{fig:placement:good}) requires only one rearrangement step.
            The atoms' current locations are depicted in grey, and their new locations in blue.
        }
        \label{fig:placement}
    \end{figure}

    \begin{example}\label{exp:improvement}
        Consider the scenario sketched in~\cref{fig:placement}, where one atom in the entanglement zone is reused from the previous \mbox{two-qubit} gate layer, and three other atoms initially located in the storage zone will be moved to the entanglement zone.
        Mainly optimizing for accumulated movement distance yields the placement as shown in~\cref{fig:placement:bad}.
        While atoms are indeed close, this placement prohibits the parallel movement of all three atoms indicated by the grey arrows as this would violate the non-crossing constraint (cf.~\cref{sec:zoned architecture}).
        Hence, the routing stage must split the movements into two rearrangement steps.
        In contrast, a placement that already considers such constraints could generate a solution as shown in~\cref{fig:placement:good}.
        Here, all movements are compatible and it requires only one rearrangement step.
    \end{example}

    The example clearly demonstrates that the overall solution can be improved by considering the rearrangement constraints during the placement stage.
    To this end, we propose the concept of \emph{\mbox{routing-aware} placement}.

    \paragraph*{Routing-Aware Placement}
    While performing the placement stage reviewed in~\cref{sec:compilation flow}, \mbox{routing-aware} placement reduces the number of rearrangement steps in the subsequent routing stage and, hence, reduces the rearrangement time during the resulting quantum computation.
    To this end, \mbox{routing-aware} placement groups placed atoms corresponding to compatible movements for each transition from one layer to the next.
    Then, the main objective is to minimize the sum of the group's rearrangement duration.
    Since all movements in each group can be performed in parallel, this objective minimizes both the number of rearrangement steps and the atoms' travel distances.

    \section[Proposed Solution: Unleashing the Potential with Routing-Aware Placement]{Proposed Solution: Unleashing the Potential with Routing-Aware Placement}\label{sec:solution}

    The previous section has shown that \mbox{routing-aware} placement could tap huge potential.
    However, the search space for possible placements is gigantic, and we need efficient methods to explore it.
    This section proposes an approach to handle the resulting complexity and, by this, utilize the potential of \mbox{routing-aware} placement.
    With this basis, \cref{sec:details} provides its implementation details before \cref{sec:evaluation} evaluates its effectiveness.

    \subsection{The Search Space}\label{sec:search space}

    The placement of the atoms is determined sequentially for every \mbox{two-qubit} gate layer and in between, resulting in two types of placements:
    \begin{itemize}
        \item A \emph{gate placement}, where all atoms involved in a \mbox{two-qubit} gate are placed in paired traps in the entanglement zone while all remaining atoms are shielded in the storage zone.
        \item An \emph{intermediate placement}, where all atoms not reused in the following \mbox{two-qubit} gate layer, are placed back in the storage zone.
    \end{itemize}
    To begin with, an \emph{initial} (intermediate) placement of the atoms in the storage zone is determined.
    Then, each placement is constructed based on the previous one.

    As illustrated in~\cref{fig:placement}, some placement solutions are more favorable than others:
    A good placement corresponds to a fast transition from the previous placement, \ie, a few rearrangement steps and short distances for the atoms to travel.
    We reflect this objective in the following cost function defined as the sum of two parts:
    \begin{enumerate*}
        \item a proxy for the actual routing cost, and
        \item a \mbox{look-ahead} part to achieve a better global solution.
    \end{enumerate*}

    \paragraph*{Proxy for Actual Cost} The exact duration of the transition is determined later by the routing stage.
    As we cannot generate an exact routing solution for each placement candidate considering its complexity~\cite{linReuseAwareCompilationZoned2024}, an efficiently computable yet accurate proxy for the routing cost is essential.
    To this end, the atom movements corresponding to the placement are grouped greedily into groups of compatible movements.
    Movements are compatible if performed in parallel without violating the rearrangement constraints reviewed in~\cref{sec:zoned architecture}.

    More precisely, let \(\operatorname{d_{\mathrm{max}}}\left(G\right)\) be the maximum travel distance of an atom in group \(G\).
    Then, the cost of a placement \(p\) is calculated as
    \vspace{-6pt}
    \begin{equation}
        \operatorname{cost}\left(p\right) \defeq \sum_{G\in\operatorname{\mathsf{groups}}\left(p\right)} \!\!\!\!\sqrt{\operatorname{d_{\mathrm{max}}}\left(G\right)}\label{eq:cost}\enspace.
    \end{equation}

    Note that the square root is used because the movement duration is proportional to the square root of the travel distance~\cite{bluvsteinQuantumProcessorBased2022}.
    This favors fewer groups with shorter movements.

    \paragraph*{Look-ahead Cost} There may be multiple choices with similar associated costs while placing atoms.
    However, some can be more beneficial for upcoming \mbox{two-qubit} gate layers than others.
    When placing an atom closer to the atom's next interaction partner, the resulting duration of the quantum computation can be further reduced.
    Furthermore, \mbox{look-ahead} allows us to detect situations where reuse is detrimental.

    \begin{figure}[tp]
        \centering
        \includegraphics[width=\linewidth]{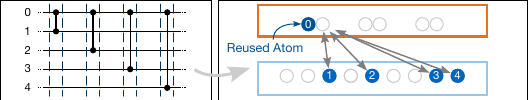}\\
        \raggedright\vspace{-16pt}\hspace{2pt}%
        \begin{subfigure}{102pt}
            \caption{\raggedright}
            \label{fig:reuse:circuit}
        \end{subfigure}
        \begin{subfigure}{100pt}
            \caption{\raggedright}
            \label{fig:reuse:placement}
        \end{subfigure}
        \vspace{-4pt}
        \caption{There are situations where reuse is detrimental to the duration.}
        \label{fig:reuse}
        \vspace{-10pt}
    \end{figure}

    \begin{example}\label{exp:reuse}
        Consider the circuit in~\cref{fig:reuse:circuit}.
        The atom~0 can be reused for all four layers.
        However, this leads to very long movements for the atoms~3 and~4, and \emph{not} reusing the atom~0 after the first two gates would be more beneficial.
    \end{example}

    Using \mbox{look-ahead}, the cost of moving the next interaction partner of a reused atom can be considered when deciding whether to reuse that atom.
    Without the \mbox{look-ahead}, the future cost of reusing an atom cannot be estimated.

    All considerations from above yield a gigantic number of possible placements---more precisely, exponentially many in the number of atoms.
    Overall, the objective is to find the best one, \ie, the one with the lowest cost.

    \begin{example}\label{exp:search space}
        Assume a modest number of eight CZ-gates were performed in parallel.
        Afterward, 16 atoms in the entanglement zone must be returned to the storage zone.
        Even if we limit the search to, \eg, the 36 nearest traps in the storage zone for each atom (cf. pruning strategies in~\cref{sec:pruning}), this results in \(36^{16} \approx 1.3 \times 10^{24}\) possible placements.
    \end{example}

    Obviously, naively enumerating all these placements is infeasible for relevant circuit sizes.
    Hence, sophisticated methods are needed to consider as few as possible placements until a \mbox{(near-)optimal} solution is found.

    \subsection{Taming the Search Space with A*}\label{sec:a*}

    To tackle this complexity, we propose to use the A*~algorithm~\cite{hartFormalBasisHeuristic1968}.
    It has already proved successful in similar scenarios to cope with a huge search space, such as routing qubits on superconducting chips~\cite{zulehnerEfficientMethodologyMapping2019}.
    To this end, finding the best placement for each layer is encoded into a \mbox{state-space} search in the following manner.
    First, all atoms that must be rearranged are identified: For a gate placement, these are the atoms involved in a gate but not reused from the \emph{previous} layer; conversely, for an intermediate placement, these are the atoms in the entanglement zone that are not reused in the \emph{next} layer.
    Second, the placement is determined following a \mbox{gate-by-gate} strategy for gate placements and an \mbox{atom-by-atom} strategy for intermediate placements.
    Note that the placement of a gate is the mapping of the respective atoms to the paired traps in the entanglement zone.
    The location of the remaining atoms is copied from the previous placement.

    Starting from a start node corresponding to the state where none of the atoms to be moved are placed yet, A* finds the path to the goal node representing the cheapest placement concerning the cost function from~\cref{sec:search space}.
    On that path, the nodes between the start and goal nodes represent intermediate partial placements.
    More precisely, a node inherits its predecessor's placement and adds one additional atom or gate.

    Since we follow an \mbox{atom-by-atom} or \mbox{gate-by-gate} strategy, the neighbors of the start node represent the possible placements of the first atom or gate, respectively.
    In turn, their neighbors correspond to the possible placements of the first two atoms or gates.
    This goes on until all atoms or gates are placed.
    This eventually results in a representation of the search space as a tree.

    \begin{figure}[tp]
        \centering
        \includegraphics[width=\linewidth]{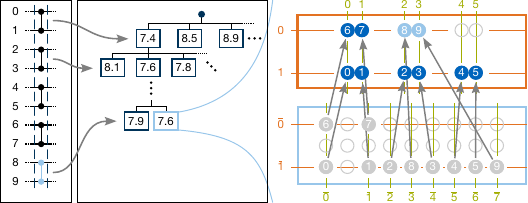}\\
        \raggedright\vspace{-15.5pt}\hspace{1pt}%
        \begin{subfigure}{28pt}
            \caption{\raggedright}
            \label{fig:method:layer}
        \end{subfigure}
        \hspace{2pt}
        \begin{subfigure}{86pt}
            \caption{\raggedright}
            \label{fig:method:search tree}
        \end{subfigure}
        \hspace{2pt}
        \begin{subfigure}{90pt}
            \caption{\raggedright}
            \label{fig:method:partial placement}
        \end{subfigure}
        \vspace{-4pt}
        \caption{
            (\subref{fig:method:layer}) a layer of five CZ gates.
            (\subref{fig:method:search tree}) the search tree for the placement of five CZ gates with the associated cost in each node.
            (\subref{fig:method:partial placement}) the placement of the atoms of the first four CZ gates (blue) and the ones of the last (light blue).
        }
        \label{fig:method}
        \vspace{-10pt}
    \end{figure}

    \begin{example}\label{exp:tree}
        Consider the \mbox{two-qubit} gate layer in~\cref{fig:method:layer} with five gates.
        The rectangular nodes in the first level of~\cref{fig:method:search tree} represent the various options to place the first gate involving atom~0 and~1 together with their estimated cost inscribed in the nodes.
        In the second level, some placement options for the second gate are shown, extending the first placement option for the first gate.
        Finally, the two nodes in the last level correspond to the placement options of the fifth gate.
        All other nodes are omitted.
        \Cref{fig:method:partial placement} shows the placement represented by the light blue node in~\cref{fig:method:search tree}, where the atoms~8 and~9 are placed in the middle of the top row.
        Since it is a node of the last level, the location of the atoms~0 to~7 is already fixed.
    \end{example}

    The A* algorithm explores the search space from the start node until it finds a goal node.
    To this end, it extends the \mbox{well-known} Dijkstra's algorithm~\cite{cormenIntroductionAlgorithms2022} by using a heuristic function to guide the search faster towards some goal.
    For every new node it encounters, it calculates the cost of the current node and adds an estimate of the remaining cost for the goal node.
    The next node expanded is the one with the lowest estimated overall cost.
    A heuristic that never overestimates the cost to reach the goal is called \emph{admissible}.
    Only with an admissible heuristic A* is guaranteed to find the optimal path.

    \subsection{The Heuristic Guiding the Search}\label{sec:heuristic}
    The sheer size of the search space quickly renders an unguided search infeasible.
    Hence, we need a good heuristic to guide the search faster toward a goal.
    The proposed heuristic is defined as the sum of three parts:
    \begin{enumerate*}[label=(\arabic*)]
        \item an admissible part,
        \item an accelerating part, and
        \item a \mbox{look-ahead} part.
    \end{enumerate*}

    \paragraph*{Admissible Part} The admissible part of the heuristic is equal to the lower bound of the cost's increase until a goal node is reached.
    For the lower bound, we assume that all remaining movements can be added to existing groups of compatible movements.
    Hence, the number of groups does not change; only the maximum distance per group may increase.

    \paragraph*{Accelerating Part} Using only the admissible part of the heuristic does not sufficiently accelerate the search.
    Whenever the actual cost is significantly higher than the estimate by the heuristic, A* goes a lot more into breadth instead of depth---slowing down the search.
    Thus, we add an accelerating part to the heuristic to guide the search more aggressively---accepting that the resulting heuristic may not be admissible anymore.
    The accelerating part estimates how likely conflicts arise when an intermediate node's partial placement is extended to a full placement.
    It, then, favors partial placement with fewer anticipated conflicts.
    To this end, we follow the following rationale: Rearrangements that retain existing gaps from the previous placement are more straightforward to extend with new placements.

    \begin{figure}[bp]
        \vspace{-8pt}
        \centering
        \includegraphics[width=\linewidth]{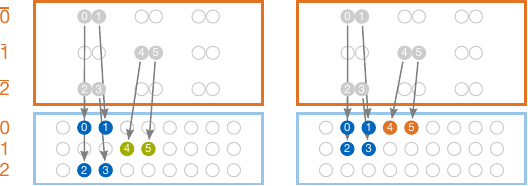}\\
        \vspace{-18pt}\hspace{9pt}%
        \begin{subfigure}{122pt}
            \caption{\raggedright}
            \label{fig:deviation:good}
        \end{subfigure}
        \begin{subfigure}{100pt}
            \caption{\raggedright}
            \label{fig:deviation:bad}
        \end{subfigure}\vspace{-4pt}
        \caption{
            (\subref{fig:deviation:good}) placement requiring one rearrangement step.
            (\subref{fig:deviation:bad}) placement requiring two rearrangement steps.
        }
        \label{fig:deviation}
    \end{figure}

    \begin{example}\label{exp:deviation}
        Consider the situation shown in~\cref{fig:deviation}, where the previous location of all atoms in the upper zone is depicted in grey.
        \Cref{fig:deviation:good} depicts a placement of the atoms~0 to~3 where the vertical gap between them is preserved.
        Consequently, the atoms~4 and~5 can be placed in the gap without causing a conflict.
        Conversely, \cref{fig:deviation:bad} shows a placement of the atoms~0 to~3 where the gap is not preserved.
        Thus, any placement of the atoms~4 and~5 causes a conflict.
    \end{example}

    To favor placements like the one in~\cref{fig:deviation:good}, the heuristic analyzes the partial placement, such as the one of the atoms~0 to~3 in~\cref{exp:deviation} and assigns a lower estimate of the additional cost to those with a similar spacing between the placed atoms as in the previous placement.

    \paragraph*{Look-ahead Part}
    After the \mbox{look-ahead} cost from \cref{sec:search space} is added to the actual cost, its value increases and outweighs the present heuristic cost, so the heuristic cost fails to accelerate the search.
    In particular, this effect diminishes the indented acceleration by the accelerating part of the heuristic.
    Thus, we also design a heuristic cost for the \mbox{look-ahead} part to recover the acceleration.
    To this end, we add the average \mbox{look-ahead} cost of all options of each unplaced atom or gate.

    The evaluation in~\cref{sec:evaluation} shows that the proposed heuristic manages to guide the search towards a solution with few rearrangement steps.
    At the same time, the increase in the placement time remains moderate because of the strategy proposed above and its following efficient implementation.

    \section{Implementation}\label{sec:details}
    Using the concepts introduced in~\cref{sec:solution}, their efficient implementation is key to the success of the proposed approach.
    In fact, to implement the proposed approach we employ a dedicated data structure to enable efficient computation of the cost and heuristic function.
    Moreover, we also incorporate pruning strategies to further reduce placement time.
    This section provides details on their implementation.

    \subsection{Dedicated Data-Structure}\label{sec:details:data structure}

    To facilitate performance, each node stores the corresponding groups of compatible movements in a dedicated data structure that allows for an efficient check of whether a new movement is compatible with the existing groups.
    To relate the relative location of the atoms in the source and target area, the source locations of the atoms are rearranged, and the target traps are discretized among all the atoms to be moved.

    \begin{example}\label{exp:discretization}
        The source location of the (grey) atom 0 in the lower left of~\cref{fig:method} is not identified by the actual row and column of the zone but by the row~\(\bar{1}\) and column~\(\bar{0}\), which is the discrete row and column among all atoms that are rearranged.
    \end{example}

    Two mappings can express every atom's movement: Its source row and column to its target row and column, respectively.
    Each group stores the row and column of the atoms with compatible moves in separate binary trees, where the entries are sorted by their source row or column, respectively.

    \begin{figure}[bp]
        \vspace{-8pt}
        \centering
        \includegraphics[width=\linewidth]{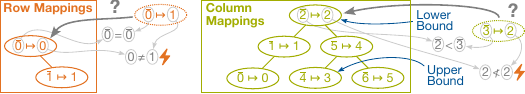}\\
        \raggedright\vspace{-16pt}\hspace{2pt}%
        \begin{subfigure}{93pt}
            \caption{\raggedright}
            \label{fig:tree:row}
        \end{subfigure}
        \begin{subfigure}{86pt}
            \caption{\raggedright}
            \label{fig:tree:column}
        \end{subfigure}
        \vspace{-4pt}
        \caption{Binary search tree containing the row and column mappings of the placement of the atoms belonging to the first four CZ-gates in~\cref{fig:method:partial placement}.}
        \label{fig:tree}
    \end{figure}

    \begin{example}\label{exp:binary tree}
        \Cref{fig:tree} shows the binary search trees containing the row and column mappings of the placement of the atoms belonging to the first four CZ-gates in~\cref{fig:method:partial placement}.
    \end{example}

    A new placement is compatible with an existing group if the new row and column mappings are compatible with the existing ones in the respective trees.
    A new mapping \(k\mapsto v\) is compatible if and only if either of the following holds:
    \begin{enumerate}
        \item\label{itm:group:exists} The key \(k\) is already contained in the tree and maps to the same value \(v\).
        \item\label{itm:group:bounds} If the key \(k\) is not contained, let \(k_{\mathrm{lower}}, k_{\mathrm{upper}}\) be the next lower and upper key mapping to \(v_{\mathrm{lower}}, v_{\mathrm{upper}}\), respectively.
            Then the new mapping is compatible if and only if \mbox{\(v_{\mathrm{lower}} < v < v_{\mathrm{upper}}\)}.
    \end{enumerate}

    The condition~\ref{itm:group:exists} ensures that all atoms in the same row or column remain during a rearrangement in the same row or column, respectively.
    The condition~\ref{itm:group:bounds} ensures that the relative order of the atoms is preserved and that no rows or columns are merged during a rearrangement.

    \begin{example}
        The placement of atom 8 in~\cref{fig:method} is not compatible with the existing placements in two ways: First, when the key \(\bar{0}\) is looked up in the row mappings depicted in~\cref{fig:tree:row}, the mapping \(\bar{0}\mapsto 0\) is found and \(0 = 1\) is not satisfied;
        Second, the column mappings shown in~\cref{fig:tree:column} do not contain the key~\(\bar{3}\).
        Hence, the next lower and upper keys are retrieved, which correspond to the mappings \(\bar{2}\mapsto 2\) and \(\bar{4}\mapsto 3\).
        The inequality \mbox{\(2 < 2 < 3\)} obviously does not hold.
    \end{example}

    When expanding a new node during the search, the respective movement of the newly placed atom is either put into an existing group of compatible movements or if no such group exists, a new group is created with the new movement.

    \subsection{The Look-Ahead}\label{sec:details:lookahead}

    The \mbox{look-ahead} cost introduced in~\cref{sec:search space} differs for gate and intermediate placements.
    For gate placements, if atom~\(a_1\) of gate~\(g_1\) is reused in the next layer by gate~\(g_2\) acting on atom~\(a_2\) and the reused atom~\(a_1\), then the \mbox{look-ahead} cost is the cost to move the atom \(a_2\) from its current location next to the reused atom \(a_1\).
    Let \(\operatorname{cost}_\mathrm{l}(g)\) be that cost for any gate \(g\) acting on a reused atom and 0 otherwise.

    For intermediate placements, if, in the next layer, gate~\(g_1\) involves atoms~\(a_1\) and~\(a_2\), atom~\(a_1\) might be identified as reusable, cf.~\cref{sec:compilation flow}.
    Then, the reuse of atom~\(a_1\) is treated as another option, just like every other placement of atom~\(a_1\).
    Hence, two cases must be distinguished: If atom~\(a_1\) is reused, it remains at its location in the entanglement zone, and the \mbox{look-ahead} cost is the cost to move atom~\(a_2\) from its current location next to the reused atom~\(a_1\).
    In this case, we subtract a constant \(\gamma\) from the \mbox{look-ahead} cost of atom~\(a_1\) corresponding to the gained fidelity by saving two trap transfers when reusing the atom.
    This cost is denoted by \(\operatorname{cost}_\mathrm{r}(a)\) for any atom \(a\).
    Conversely, if atom~\(a_1\) is not reused or not reusable in the first place, it is moved to the storage zone and the \mbox{look-ahead} cost is equal to the cost to move atom~\(a_2\) from its current location to the new location of atom~\(a_1\).
    Let this cost be denoted by \(\operatorname{cost}_\mathrm{l}(a)\) for any atom \(a\) if in the next layer a gate acts on atom~\(a\) and it is not reused; otherwise, \(\operatorname{cost}_\mathrm{l}(a)\) is 0.

    In the following formal definition of the final cost function of a (partial) placement \(p\), the user-defined parameter \(\alpha\) adjusts the influence of the \mbox{look-ahead}.
    The cost for reusing an atom is purposefully not affected by \(\alpha\) because it represents the actual cost of the movement of its interaction partner in the next layer rather than being an estimate for future costs.
    Overall, this yields
    \vspace{-6pt}
    \begin{equation}
        \operatorname{cost}^*\left(p\right) \defeq \operatorname{cost}\left(p\right) + \!\!\!\!\!\!\!\!\sum_{a \in\set{\substack{\text{placed reused}\\\text{atoms}}}\!\!\!\!\!\!}\!\!\!\!\!\!\!\!\!\operatorname{cost}_\mathrm{r}(a) + \alpha\cdot \!\!\!\!\!\!\!\!\!\sum_{a \in\set{\substack{\text{placed}\\\text{gates/atoms}}}\!\!\!\!}\!\!\!\!\!\!\!\!\operatorname{cost}_\mathrm{l}(a) \label{eq:cost function}\enspace.
    \end{equation}

    \subsection{The Heuristic}\label{sec:deteils:heuristic}

    As stated in~\cref{sec:heuristic}, the admissible part only covers the additional cost for the goal node in the optimal case, \ie, when all future placements are compatible with existing ones.
    In this case, the number of summands in~\cref{eq:cost} remains fixed, and the cost may only increase when the maximum distance of a group increases.
    The maximum distance of any group is bound from below by the maximum distance of an unplaced atom to its nearest free trap.
    The increase of the cost function, which uses the square root of the distance, is then bound from below by the following difference defining the admissible part of the heuristic
    \vspace{-6pt}
    \begin{equation}
        \operatorname{h}\left(p\right) \defeq \max_{\substack{G\in\\\operatorname{\mathsf{groups}}\left(p\right)}} \!\!\sqrt{\operatorname{d_{\mathrm{max}}}\!\left(G\right)} \;-\! \max_{\substack{a\in\\\set{\text{unplaced atoms}}\!\!\!\!\!\!\!\!}} \!\!\!\!\!\!\sqrt{\operatorname{d_{min}}(a)}\enspace.
    \end{equation}

    To return an estimate of the likelihood of conflicts, the heuristic's accelerating part first calculates the difference between every value and key in each binary tree.
    Then, it computes the standard deviation of these differences per binary tree representing the groups of compatible movements.
    Those standard deviations are then summed up and multiplied by the number of atoms that still need to be placed because a higher deviation is more problematic when more atoms must still be placed.
    Before the multiplication, a constant \(\beta\) is added to the sum to prioritize nodes further down in the tree.
    The result is multiplied by a user-defined factor \(\delta\) to adjust the influence of this factor and added to the previous heuristic, leading to
    \vspace{-6pt}
    \begin{equation}
        \underline{\operatorname{h}}\left(p\right) \defeq \operatorname{h}\left(p\right) + \delta \cdot \left(\beta + \sum_{G\in\operatorname{\mathsf{groups}}\left(p\right)\!\!\!\!\!\!\!\!\!\!\!\!}\!\!\operatorname{\mathsf{SD}}\left(G\right)\right)\cdot\abs{\set{\substack{\text{unplaced}\\\text{gates/atoms}}}}\enspace,
    \end{equation}
    where \(\operatorname{\mathsf{SD}}\) is the standard deviation of the key-value differences.

    To handle significant discrepancies in column and row count between the discrete source and target locations, the key is multiplied with a scaling factor before subtracting it from the value.
    This factor is determined such that a rearrangement with a standard deviation close to 0 corresponds to a more or less parallel movement of atoms without changing the spacing between them.

    \begin{figure}[bp]
        \vspace{-8pt}
        \centering
        \includegraphics[width=\linewidth]{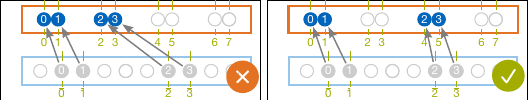}\\
        \raggedright\vspace{-16pt}\hspace{2pt}%
        \begin{subfigure}{125pt}
            \caption{\raggedright}
            \label{fig:scaling:without}
        \end{subfigure}
        \begin{subfigure}{80pt}
            \caption{\raggedright}
            \label{fig:scaling:with}
        \end{subfigure}
        \vspace{-4pt}
        \caption{By scaling keys before calculating the standard deviation the heuristic favors parallel movements, cf.~\cref{exp:scaling}.}
        \label{fig:scaling}
    \end{figure}

    \begin{example}\label{exp:scaling}
        \Cref{fig:scaling:without} shows a placement that the heuristic would favor without the scaling factor because all key-value differences are equal and the standard deviation is zero.
        By multiplying the key with the scaling factor \(2\) (8 target cols. divided by 4 source cols.), the heuristic prefers the better placement in~\cref{fig:scaling:with} instead.
        Here, the standard deviation of the differences \([0, -1, 0, -1]\) including the scaling is \(0.5\) compared to, \eg, \(1.1\) for the placement in~\cref{fig:scaling:without}.
    \end{example}

    As mentioned in~\cref{sec:heuristic}, the heuristic's \mbox{look-ahead} part adds the average \mbox{look-ahead} costs across all options to keep the accelerating effect.
    Formally, the variable \(\widebar{\mathsf{cost}}_\mathrm{l}(a)\) represents the average \mbox{look-ahead} across all placement options of atom or gate \(a\) (this includes the full costs for reused atoms and the with \(\alpha\) scaled \mbox{look-ahead} costs).
    This results in the final heuristic function
    \vspace{-6pt}
    \begin{equation}
        \underline{\operatorname{h}}^*\left(p\right) \defeq \underline{\operatorname{h}}\left(p\right) + \sum_{a \in\set{\substack{\text{unplaced}\\\text{gates/atoms}}}} \widebar{\mathsf{cost}}_\mathrm{l}(a) \label{eq:heuristic}\enspace.
    \end{equation}

    \subsection{Pruning Strategies}\label{sec:pruning}

    Only the most likely target traps are considered during the search to reduce the number of potential nodes in the search space.
    To this end, an adjustable-size window is employed and centered around the nearest target trap for each atom.
    All target traps within this window are considered placement options during the search.
    Should the case arise that the window is too small in the sense that it contains too few free traps, the window is expanded automatically.

    \begin{table*}[tp]
        \setlength{\aboverulesep}{0.4pt} %
        \setlength{\belowrulesep}{0.4pt} %
        \caption{Routing-Agnostic vs. Routing-Aware Placement}
        \vspace{-6pt}
        \label{tab:results}
        \begin{tabularx}{\linewidth}{%
            X%
            S[table-format=3.0]%
            S[table-format=5.0]%
            S[table-format=4.0]%
            S[table-format=2.0]%
            |S[table-format=4.1]%
            S[table-format=2.2]%
            S[table-format=4.0]%
            S[table-format=4.1]%
            |S[table-format=4.1]%
            S[table-format=2.2]%
            S[table-format=4.0]%
            @{\ (\hspace{.2ex}}r@{)\hspace{2ex}} %
            S[table-format=4.1]%
            @{\ (\hspace{.2ex}}r@{)\hspace{2ex}} %
        }
            \toprule
            \multicolumn{4}{l}{\textbf{Benchmark}} & {Max.} & \multicolumn{4}{c|}{\textbf{Routing-Agnostic} \textit{(State-of-the-Art)}} & \multicolumn{6}{c}{\textbf{Routing-Aware} \textit{(Proposed Solution)}} \\[-1pt] %
            \cmidrule{6-9}\cmidrule{10-15}
            \multicolumn{2}{r}{Num.} & {Num.} & {Num.} & {\hspace{-3pt}2Q-Gates\hspace{-3pt}} & \multicolumn{2}{c}{Time [\si{\milli\second}]} & {Num. R.} & {Rearr.} & \multicolumn{2}{c}{Time [\si{\milli\second}]} & \multicolumn{2}{c}{Num. Rearr.} & \multicolumn{2}{c}{Rearr.} \\[-1pt] %
            \cmidrule(r){6-7}\cmidrule(r){10-11}
            \multicolumn{2}{r}{Qubits} & {2Q-Gates} & {Layers} & {\hspace{-3pt}in Layer\hspace{-3pt}} & {Place.} & {Rout.} &  {Steps} & {T. [\si{\milli\second}]}  & {Place.} & {Rout.} & \multicolumn{2}{c}{Steps} & \multicolumn{2}{c}{Time [\si{\milli\second}]} \\ %
            \midrule
            graphstate\textsuperscript{1} &  60 &    60 &     4 & 22 &    3.3 &    0.15 &    56 &    8.5 &   27.4 &   0.13 &    37 & \bfseries --34\% &    6.9 & \bfseries --19\% \\      %
                                          & 100 &   100 &     6 & 37 &   10.4 &    0.27 &    88 &   17.3 & 1860.5 &   0.30 &    62 & \bfseries --30\% &   11.9 & \bfseries --31\% \\[2pt] %
            wstate\textsuperscript{1}     & 200 &   398 &   201 &  2 &   21.6 &    0.23 &   643 &   83.5 &   98.4 &   0.21 &   493 & \bfseries --23\% &   57.4 & \bfseries --31\% \\      %
            qft\textsuperscript{1}        & 200 &  9350 &   794 & 13 &  374.0 &    6.32 &  4300 &  619.9 & 1309.3 &   4.75 &  3534 & \bfseries --18\% &  546.8 & \bfseries --12\% \\      %
                                          & 500 & 24350 &  1994 & 13 & 1138.4 &   15.61 & 10391 & 1635.7 & 3060.8 &  14.10 &  9911 & \bfseries  --5\% & 1554.9 & \bfseries  --5\% \\[2pt] %
            qpeexact\textsuperscript{1}   & 200 &  9697 &  1187 & 13 &  489.6 &    5.64 &  4766 &  692.8 & 1482.7 &   5.13 &  4306 & \bfseries --10\% &  651.8 & \bfseries  --6\% \\[2pt] %
            ising\textsuperscript{2}      &  42 &    82 &     4 & 21 &    2.4 &    0.07 &    22 &    3.1 &  451.2 &   0.05 &     9 & \bfseries --59\% &    1.6 & \bfseries --49\% \\      %
                                          &  98 &   194 &     4 & 49 &   16.9 &    0.23 &    23 &    3.2 &  417.9 &   0.16 &    12 & \bfseries --48\% &    1.9 & \bfseries --43\% \\[2pt] %
            qft\textsuperscript{2}        &  18 &   306 &    66 &  9 &    8.7 &    0.13 &   185 &   21.9 &   32.7 &   0.10 &   148 & \bfseries --20\% &   20.3 & \bfseries  --8\% \\      %
                                          &  29 &   680 &   110 &  9 &   18.6 &    0.36 &   434 &   57.6 &   64.4 &   0.21 &   284 & \bfseries --35\% &   40.5 & \bfseries --30\% \\[2pt] %
            bv\textsuperscript{2}         &  30 &    18 &    18 &  1 &    1.1 &    0.01 &    37 &    4.3 &    5.4 &   0.01 &    36 & \bfseries  --3\% &    4.1 & \bfseries  --3\% \\      %
                                          &  70 &    36 &    36 &  1 &    3.3 &    0.01 &    73 &   11.0 &   11.4 &   0.01 &    76 &             +4\% &    8.5 & \bfseries --23\% \\[2pt] %
            wstate\textsuperscript{2}     &  27 &    52 &    28 &  2 &    2.4 &    0.02 &    73 &    8.3 &   11.1 &   0.02 &    62 & \bfseries --15\% &    6.7 & \bfseries --19\% \\      %
            seca\textsuperscript{2}       &  11 &    80 &    37 &  3 &    3.0 &    0.03 &   112 &   12.2 &   13.2 &   0.02 &    81 & \bfseries --28\% &    9.9 & \bfseries --18\% \\      %
            ghz\textsuperscript{2}        &  40 &    39 &    39 &  1 &    2.1 &    0.01 &    78 &    9.6 &   12.4 &   0.01 &    79 &             +1\% &    8.7 & \bfseries --10\% \\      %
                                          &  78 &    77 &    77 &  1 &    6.0 &    0.03 &   154 &   20.9 &   23.5 &   0.03 &   157 &             +2\% &   17.4 & \bfseries --17\% \\[2pt] %
            multiply\textsuperscript{2}   &  13 &    40 &    23 &  3 &    1.6 &    0.02 &    68 &    7.6 &    8.1 &   0.02 &    54 & \bfseries --21\% &    6.5 & \bfseries --15\% \\      %
            cat\textsuperscript{2}        &  22 &    21 &    21 &  1 &    0.9 &    0.01 &    43 &    4.6 &    6.5 &   0.01 &    42 & \bfseries  --2\% &    4.5 & \bfseries  --2\% \\      %
                                          &  35 &    34 &    34 &  1 &    1.9 &    0.01 &    69 &    8.2 &   10.4 &   0.01 &    69 &      \textpm 0\% &    7.5 & \bfseries  --8\% \\[2pt] %
            swaptest\textsuperscript{2}   &  25 &    84 &    62 & 12 &    4.3 &    0.05 &   127 &   13.4 &   27.3 &   0.05 &   121 & \bfseries  --5\% &   12.3 & \bfseries  --8\% \\      %
            knn\textsuperscript{2}        &  31 &   105 &    77 & 15 &    5.4 &    0.08 &   158 &   16.7 &   63.1 &   0.06 &   149 & \bfseries  --6\% &   15.5 & \bfseries  --7\% \\      %
            \midrule
            \(\varnothing\)               &     &       &       &    &  100.8 &    1.40 &   155 & 1042.9 &  428.5 &   1.21 &   143 & \bfseries --17\% &  939.1 & \bfseries --17\% \\      %
            \bottomrule
        \end{tabularx}
        \vspace{-12pt}
    \end{table*}

    \section{Evaluation}\label{sec:evaluation}

    The proposed method constitutes the first \mbox{routing-aware} placement for zoned neutral atom architectures.
    To evaluate its effectiveness, all methods and data structures have been implemented and, afterward, compared to the \mbox{state-of-the-art} \mbox{routing-agnostic} placement method.
    This section summarizes the correspondingly obtained results.
    To this end, we first describe the setup and results of a parameter study before the results of this comparison are provided and discussed.

    \subsection{Experimental Setup and Parameter Study}\label{sec:setting}

    We implemented the proposed method in C++.
    To demonstrate its effectiveness, we afterward compared it against the \mbox{state-of-the-art} compiler ZAC~\cite{linReuseAwareCompilationZoned2024}.
    For a fair runtime comparison, we re-implemented ZAC, including the \mbox{routing-agnostic} placement in C++.
    The complete code is publicly available in open-source as part of the MQT~\cite{willeMQTHandbookSummary2024} under \url{https://github.com/munich-quantum-toolkit/qmap}.
    All experiments were conducted on an Apple M3 with 16GB of RAM.

    As benchmark circuits, we used the ones considered originally for the evaluation of ZAC~\cite{linReuseAwareCompilationZoned2024} (taken from \mbox{QASMBench}~\cite{li2023qasmbench}).
    This set is complemented with additional circuits of larger sizes (taken from MQT~Bench~\cite{quetschlichMQTBenchBenchmarking2023}).

    The \mbox{routing-aware} placement proposed above offers several parameters to tune its performance.
    We conducted a parameter study to find suitable parameter combinations for each of the two benchmark sets.
    To this end, we varied the parameters \(\alpha\), \(\beta\), \(\gamma\), and \(\delta\) and collected the number of rearrangement steps, the rearrangement time, and the time for the placement stage.
    This allows us to determine proper parameter settings.

    \begin{figure}[tp]
        \centering
        \begin{subfigure}{30mm}
            \centering
            \includegraphics[height=27mm]{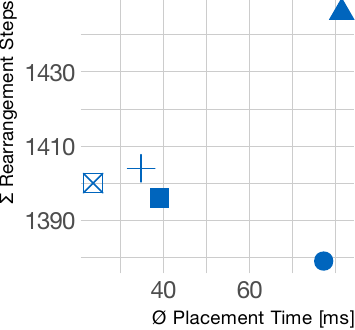}\\
            \vspace{-16pt}
            \caption{\raggedright}
            \label{fig:ablation study:steps}
        \end{subfigure}
        \begin{subfigure}{57mm}
            \centering
            \includegraphics[height=27mm]{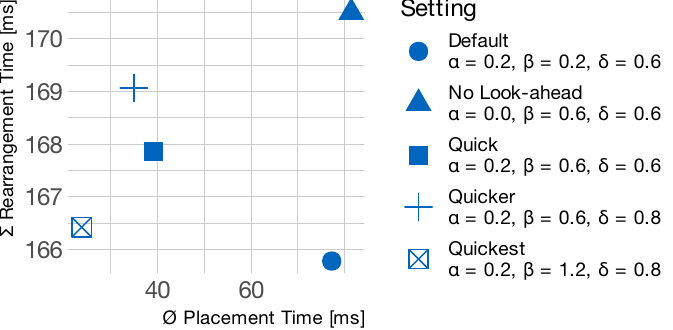}\\
            \vspace{-16pt}
            \caption{\raggedright}
            \label{fig:ablation study:time}
        \end{subfigure}
        \vspace{-16pt}
        \caption{Trade-off between placement time and result quality.}
        \label{fig:ablation study}
        \vspace{-10pt}
    \end{figure}

    As an example, \cref{fig:ablation study} shows the results of this parameter study for the \mbox{QASMBench} benchmark set.\footnote{
        Due to space limitations, we skipped listing corresponding results for the MQT Bench benchmark set.
    }
    More precisely, \cref{fig:ablation study:steps} shows the sum of the rearrangement steps overall benchmarks in the QASMBench set against the average time for the placement stage for five different parameter combinations.
    \Cref{fig:ablation study:time}, then, shows the corresponding accumulated rearrangement time.
    Since the goal is to minimize the rearrangement steps and time, the parameter combination corresponding to the lowest values in the plots yields the highest-quality solutions.
    In particular, it shows that the \mbox{look-ahead} is crucial in improving the quality of results.

    Based on that, we derived that the parameter combination \(\alpha = 0.2\), \(\beta = 0.2\), \(\gamma = 5\), and \(\delta = 0.6\) is best for the benchmark set from \mbox{QASMBench}.
    Analogously, the larger circuits from MQT Bench, the parameter combination \(\alpha = 0.2\), \(\beta = 0.8\), \(\gamma = 5\), and \(\delta = 0.9\) led to the best results.

    \subsection{Comparison to the State-of-the-Art}\label{sec:results}

    Using the setup described above, we eventually compared the performance of the proposed approach against the \mbox{state-of-the-art}, \ie, the approach proposed in~\cite{linReuseAwareCompilationZoned2024}.
    To this end, we considered the following metrics:
    \begin{itemize}
        \item \emph{Placement time}: The time required by the placement stage.
        \item \emph{Routing time}: The time required by the routing stage.
        \item \emph{Rearrangement steps}: The number of rearrangement steps required during the execution of the quantum computation based on the compilation result.
        \item \emph{Rearrangement time}: the total time required to rearrange the atoms during the quantum computation based on the compilation result.
        This time time is calculated based on the relation \(t = (d \,/\, \SI{2750}{\meter\per\second\squared})^{1/2}\) and \SI{15}{\micro\second} for every trap transfer~\cite{bluvsteinQuantumProcessorBased2022}.
    \end{itemize}

    The obtained results are summarized in \cref{tab:results}.
    The first section of \cref{tab:results} lists the benchmarks and summarizes key statistics relevant to the placement and routing, namely, the number of qubits, \mbox{two-qubit} gates, and layers.
    Hereby, a layer corresponds to a set of independent \mbox{two-qubit} gates part of the sequence returned by the scheduler, cf.~\cref{sec:compilation flow}.
    Additionally, we list the maximum number of \mbox{two-qubit} gates in a layer as this determines the size of the search space, cf.~\cref{sec:search space}.
    Benchmarks stemming from MQT Bench are marked with \textsuperscript{1} and those from QASMBench with \textsuperscript{2}.

    The second and third sections summarize the obtained results of the \mbox{routing-agnostic} placement (\ie, the \mbox{state-of-the-art} method proposed in~\cite{linReuseAwareCompilationZoned2024}) and the \mbox{routing-aware} placement (proposed in this work), respectively.
    For both approaches, the metrics mentioned above, \ie, placement time, routing time, number of rearrangement steps, and rearrangement time, are listed.
    Animations of the resulting atom's rearrangements for selected benchmarks are available under \url{https://doi.org/10.5281/zenodo.15236196}.

    First, the results show that the time for the placement increases due to the gigantic search space.
    This was expected (see also discussion in~\cref{sec:search space} and~\cref{exp:search space}), but, using the proposed approach and its efficient implementation, the overhead remains moderate.
    In fact, \emph{all} benchmarks (the small ones previously considered in~\cite{linReuseAwareCompilationZoned2024} but also the larger instances) can be placed in a few seconds or even a fraction of it.

    At the same time, the results clearly show that the \mbox{routing-aware} approach (and, hence, the consideration of the larger search space) is absolutely worth it: In fact, the proposed \mbox{routing-aware} approach significantly reduces the number of required rearrangement steps on most of the benchmarks, especially, those exhibiting a large degree of parallelism.
    For example, the rearrangement steps were almost halved for the \texttt{ising} benchmark with 98 qubits.
    On average, the number of rearrangement steps is reduced by 17\% across all benchmarks.

    This substantially reduces the corresponding rearrangement times (one main metric affecting the fidelity of the quantum computation).
    In fact, these times are consistently lower for \emph{all} benchmarks
    also for the cases where the rearrangement steps slightly increased.
    This is because the additional rearrangement steps reduce the travel distance and, consequently, the rearrangement time.
    Overall, the \mbox{routing-aware} placement reduces the rearrangement time by up to 49\% in the best case and 17\% on average across all benchmarks.

    \section{Conclusions}\label{sec:conclusions}

    In this work, we presented the first \mbox{routing-aware} placement for zoned neutral atom architectures.
    In contrast to existing compilers, the proposed approach considers rearrangement constraints during the placement stage.
    Evaluations demonstrated that this approach effectively reduces the number of rearrangement steps in the subsequent routing stage.
    Especially on benchmarks exhibiting a large degree of parallelism, the proposed method can almost halve the number of rearrangement steps.
    Its implementation is publicly available in open-source as part of the MQT under \url{https://github.com/munich-quantum-toolkit/qmap}.
    
    \subsection*{Acknowledgements}\label{sec:ack}

    \footnotesize
    We thank Ludwig Schmid and Lukas Burgholzer for the fruitful discussion on the heuristic function.
    This work received funding from the European Research Council (ERC) under the European Union’s Horizon 2020 research and innovation program (grant agreement No. 101001318), was part of the Munich Quantum Valley, which the Bavarian state government supports with funds from the Hightech Agenda Bayern Plus, and has been supported by the BMWK based on a decision by the German Bundestag through project QuaST, as well as by the BMK, BMDW, and the State of Upper Austria in the frame of the COMET program (managed by the FFG). Furthermore, the work was funded by NSF grant CCF-2313083.

    \clearpage
    \renewcommand*{\bibfont}{\footnotesize}
    \printbibliography
\end{document}